\documentclass{article}
\usepackage{amsmath,amssymb,graphicx,amsfonts,epsf}
\usepackage[noend]{algpseudocode}
\usepackage{algorithm}
\usepackage{tikz,xcolor}
\usepackage{graphicx} 
\usetikzlibrary{backgrounds,decorations.pathmorphing,arrows}
\usepackage{amsopn}

\def\EE{\mathbb E}

\def\cH{\mathcal H}

\def\cH{\mathcal H}

\def\cL{\mathcal L}

\def\cX{\mathcal X}

\def\bT{\mathbf T}

\def\b1{\mathbf 1}

\newcommand{\qed}{\hfill$\square$\bigskip}
\newcommand{\raf}[1]{(\ref{#1})}
\newcommand{\proof}{\noindent {\bf Proof}~~}

\newcommand{\polylog}{\operatorname{polylog}}
\newcommand{\quasipoly}{\operatorname{quasi-poly}}

\newcommand{\hide}[1]{}

\def \PKC {\textsc{Proper-$\cL$-Coloring}}

\def\np{NP}
\newtheorem{theorem}{Theorem}

\newtheorem{lemma}{Lemma}
\newtheorem{corollary}{Corollary}
\newtheorem{proposition}{Proposition}

\newtheorem{remark}{Remark}

\newtheorem{claim}{Claim}

\title{Quasi-polynomial Algorithms for List-coloring of Nearly Intersecting Hypergraphs}
\author{ 
Khaled Elbassioni\thanks {Khalifa University of Science and Technology, Abu Dhabi, UAE;
(khaled.elbassioni@ku.ac.ae)}
}

\begin{document}
\date{}
\maketitle
\begin{abstract}
A hypergraph $\cH$ on $n$ vertices and $m$ edges is said to be {\it nearly-intersecting} if every edge of $\cH$ intersects all but at most polylogarthmically many (in $m$ and $n$) other edges. Given lists of colors $\cL(v)$, for each vertex $v\in V$, $\cH$ is said to be $\cL$-(list) colorable, if each vertex can be assigned a color from its list such that no edge in $\cH$ is monochromatic. We show that list-colorability for any nearly intersecting hypergraph, and lists drawn from a set of constant size, can be checked in quasi-polynomial time in $m$ and $n$.
\end{abstract}

\section{Introduction}
\label{intro}

\textsc{Hypergraph $k$-Coloring} is the problem of checking whether the vertex-set of a given hypergraph (family of sets) can be colored with at most $k$ colors such that every edge receives at least two {\it distinct} colors. It is a basic problem in theoretical computer science and discrete mathematics which has received considerable attention~(see, e.g. \cite{BBTV08,BL02,DG13,DRS05,GHHSV14,KS14,v2002}). The problem is \np-complete already for $k=2$, and in fact, it is quasi-\np-hard\footnote{More precisely, there is no polynomial time algorithm unless \np$\subseteq$ DTIME$(2^{\polylog n})$} to decide if a 2-colorable hypergraph can be (properly) colored with $2^{(\log n)^{\Omega(1)}}$ colors \cite{GHHSV14}. On the other hand, the best positive result is polynomial time algorithms that can color an $O(1)$-colorable hypergraph with $n^{\Omega(1)}$ colors, where $n$ is the number of vertices (see, e.g., \cite{AKMH96,CF1996,KNB01}). 
Several generalizations of the problem have also been considered, for example, \textsc{ List-Coloring} where every vertex can take only colors from a given list of colors \cite{ERT79,V76}. 
   
Given the intrinsic difficulty of the problem, it is natural to consider special classes of hypergraphs for which the problem is easier. Some better results exist for special classes, e.g., better approximation algorithms for hypergraphs of low discrepancy and rainbow-colorable hypergraphs \cite{BGL15}, polynomial time algorithms for bounded-degree linear hypergraphs \cite{BL02,CR07}, for random $3$-uniform $2$-colorable hypergraphs \cite{PS09}, as well as for some special classes of graphs \cite{DFGK99,GKK2002,JS97,W97}.      

\medskip

In this paper, we consider a special class of hypergraphs in which every edge intersects all but at most $c$ other edges (also considered for $c=0$ in \cite{S74}); we call such hypegraphs $c$-{\it intersecting}\footnote{It would have been more descriptive to call these hypergraphs $c$-{\it avoiding}, but we choose to call them $c$-intersecting to emphasize the ``intersection'' property.} for any $c\ge 0$, and {\it nearly intersecting} when $c$ is {\it polylogarithmic} in the number of vertices and edges (this is in contrast to \cite{BL02} which considers {\it nearly disjoint} hypergraphs). While near-intersection may seem as a strong restriction at a first thought, the problem is still actually highly non-trivial. In fact, the case $k=2$ and $c=0$ is equivalent to the well-known \textsc{Monotone Boolean Duality Testing}, which is the problem of checking for a given pair of CNF and DNF formulas if they represent the same monotone Boolean function \cite{EG95,S74}. Determining the exact complexity of this duality testing problem is an outstanding open question, which has been referenced in a number of complexity theory retrospectives, e.g.,  \cite{Lov92,P97}, and has been the subject of many papers, see, e.g., \cite{BI95,BM09,D97,E08,EG95,EGM02,EGM03,EMG06,FK96,G04,GK04,GM14,KS03-IPL,Takata02}. 
Fredman and Khachiyan \cite{FK96} gave an algorithm for solving this problem with running time $n^{o(\log n)}$, where $n$ is the size of the input, thus providing strong evidence that this decision problem is unlikely to be NP-hard. 

The reduction from \textsc{Boolean Duality Testing} to checking $2$-colorability is essentially obtained by a construction from \cite{S74} which reduces the problem to checking if a monotone Boolean function given by its CNF is {\it self-dual}. However, almost all the algorithms for solving {\textsc Boolean duality testing} cannot work directly with the self-duality (and hence $2$-colorabilty) problem, due to their recursive nature which results in subproblems that do not involve checking self-duality. 
The only algorithm we are aware of that works directly on the $2$-colorability version is the one given in \cite{GK04}, but it yields weaker bounds $n^{O(\log n)}$ than those given in \cite{FK96}. In this paper, we provide bounds that match closely those given in \cite{FK96} and show that those can be in fact extended to any constant $k$ on the more general class of $c$-intersecting hypergraphs and further for checking list-colorability.

We remark that, while any $0$-intersecting hypergraph is trivially $3$-colorable, the question becomes non-trivial for $c>0$. In particular, it is easy to see that any $c$-intersecting hypergraph is a $(3+c)$-colorable, and thus the question of $k$-colorability becomes interesting for any $k$ between $2$ and $3+c$. 
It is also worth mentioning that $1$-intersecting hypergrpahs have been considered in \cite[Section 2.4.1]{P08}, where it was shown that if such a hypergraph is $2$-colorable then it is also list colorable for any lists of size $2$. It is not clear whether such result extends to the cases $k>2$ or $c>1$.  

\section{Basic Notation and Main Result}
Let $\cH\subseteq 2^V$ be a hypegraph on a finite set $V$, $k\ge 2$ be a positive integer, and $\cL:V\to2^{[k]}$ be a mapping that assigns to each vertex $v\in V$ a list of {\it admissible colors} $\cL(v)\subseteq[k]:=\{1,\ldots,k\}$. An {\it $\cL$-(list) coloring} of $\cH$ is an assignment $\chi:V\to[k]$ of colors to the vertices of $\cH$ such that $\chi(v)\in\cL(v)$ for all $v\in V$. An $\cL$-coloring is said to be {\it proper} if it results in no {\it monochromatic} edges, that is, if $|\chi(H)|\ge 2$, for all $H\in\cH$, where $\chi(H):=\{\chi(v):~v\in H\}$. 

%For a $\delta\in[0,1]$, we say that a hypergraph $\cH$ is {\it $\delta$-densely intersecting} if for all $S\subseteq V$:
%\begin{equation}\label{dense}   
%\deg_{\cH_S}(H)\ge \delta|\cH_S|, \text{ for all $H\in \cH_S$}.  
%\end{equation}
%For $\delta=1$, a  $\delta$-densely intersecting hypergraph is simply called intersecting.
%A hypergraph $\cH$ is said to be {\it intersecting} if
%\begin{equation}\label{intersecting}
%H\cap H\neq\emptyset\text{ for all } H,~H'\in\cH. 
%\end{equation}
\medskip

For a non-negative integer $c$, a hypergraph $\cH$ is said to be {\it $c$-intersecting} if for all $H\in\cH$,
\begin{equation}\label{intersecting}
H\cap H'=\emptyset\text{ for at most $c$ edges } H'\in\cH. 
\end{equation}
A hypergraph is said to be {\it nearly} interesting if it is $c$-intersecting for $c=\polylog (m,n)$. 
In this paper, we are interested in the following problem:
\begin{itemize}
	\item[] \PKC: Given a hypergraph $\cH\subseteq 2^V$ satisfying \raf{intersecting} and a mapping $\cL:V\to2^{[k]}$, either find a proper $\cL$-coloring of $\cH$, or declare that no such coloring exists. 
\end{itemize}

We denote by $n:=|V|$, $m:=|\cH|$, $\nu:=\min_{v\in V}|\cL(v)|$, $\rho:=\max_{v\in V}|\cL(v)|$, and $\kappa:=\max_{u,v\in V,~u\ne v}|\cL(u)\cap\cL(v)|$. We assume without loss of generality that $\nu\ge 2$.

\medskip

For a set $S\subseteq V$, let $H_S:=\{H\in\cH:~H\subseteq S\}$ be the subhypergraph of $\cH$ induced by set $S$, $\cH^S=\{H\cap S:~H\in\cH\}$ be the projection (or trace) of $\cH$ into $S$ (this can be a multi-subhypergraph), and $\cH(S):=\{H\in\cH:~H\cap S\ne\emptyset\}$be the subhypergraph with edges having non-empty intersection with $S$. We define further $\bar S=V\setminus S$,
% and $\deg_{\cH}(S)=|\{H\in\cH:~H\cap S\neq\emptyset\}|$. 
and for $v\in V$, $\deg_{\cH}(v):=|\{H\in\cH:~v\in H\}|$. 

\medskip

Our main result is that the problem can be solved in quasi-polynomial time\footnote{that is, the running time is bounded by $2^{\polylog(N)}$ on an instance of input size $N$.} for nearly intersecting hypergraphs and constant number of colors. In fact, we will prove the following stronger result.

\begin{theorem}\label{t1}
Problem \PKC\ can be solved in $\quasipoly(m,n)$ time if $k=O(1)$ and $c=\polylog(m,n)$, with an algorithm whose recursion-tree depth is $\polylog(m,n)$. 
\end{theorem}
As a corollary, we obtain the following result on the parallel complexity of the problem (in the PRAM model).

\begin{corollary}\label{c1}
	Problem \PKC\ can be solved in $\polylog(m,n)$ parallel time on $\quasipoly(m,n)$ number of processors, if $k=O(1)$ and $c=\polylog(m,n)$. 
\end{corollary}

In the following, we will consider partial $\cL$-colorings $\chi:V\to[0:k]:=\{0,1,\ldots,k\}$ of $\cH$, where $\chi(v)=0$ is used to mean that the vertex $v$ is not assigned any color yet; we say that such coloring is proper if no (fully colored) edge is monochromatic. Given a proper partial $\cL$-coloring $\chi$ of a hypergraph $\cH\subseteq 2^V$, we will use the following notation:  
$V_0(\chi):=\{v\in V:~\chi(v)=0\}$ and $\cH_i(\chi):=\{H\in\cH:~\chi(H)=\{0,i\}\}$ for $i\in[0:k],$ and shall simply write $V_0$ and $\cH_i$ when $\chi$ is clear from the context; any extension of $\chi$ (obtained by coloring some vertices in $V_0$) will be called proper, if it results in no monochromatic edge (that is, when combined with $\chi$, it yields a proper partial $\cL$-coloring for $\cH$); $\cH(\chi)$ denotes the hypergraph $\cH$ after deleting monochromatic edges, that is, $\cH(\chi):=\cH\setminus\{H\in\cH:~|\chi(H)|\ge 2\}$. For $i\in[0:k]$, we write $\bar{\cH}_i:=\bigcup_{j\ne i}\cH_j$. 
%For a set $S\subseteq V$, we denote by $\chi[S]$ the restriction of $\chi$ on $S$. 
For two (partial) $\cL$-colorings $\chi:V\to[0:k]$ and $\chi':S\to[k]$, where $\chi(S)=\{0\}$, we denote by $\chi'':=\chi\cup\chi':V\to[0:k]$ the partial $\cL$-coloring that assigns $\chi''(v):=\chi(v)$ for $v\in V\setminus S$ and $\chi''(v):=\chi'(v)$ for $v\in S$. If there is an $H\in\cH$ such that $|H|\le 1$, we shall assume that $\cH$ is {\it not} properly $\cL$-colorable for any $\cL:V\to2^{[k]}$. Also, by assumption, an empty hypergraph (that is, $\cH=\emptyset$) is properly $\cL$-colorable. 

\medskip

Given a proper partial $\cL$-coloring $\chi$ of $\cH$, we call $\chi_0:V_0\to[k]$ a {$0$-simple} (resp., $i$-simple, for $i\in[k]$) assignment if it is obtained by choosing, for each $H\in\cH_0$ (resp., for each $H\in\cH_i$), two distinct vertices $v,v'\in H\cap V_0$ and two distinct colors $i\in\cL(v)$ and $j\in\cL(v')$ (resp., a vertex $v\in H\cap V_0$ and a color for $v$ among the colors in $\cL(v)\setminus \{i\}$). Such an assignment is proper, if the coloring $\chi\cup\chi_0$ is a proper partial $\cL$-coloring for $\cH$. The number of $0$-simple (resp., $i$-simple, for $i\in[k]$) assignments is at most $(|V_0|\rho)^{2|\cH_0|}$ (resp., $(|V_0|\rho)^{|\cH_i|}$).

\medskip

In the following two sections we give two algorithms for solving the problem. They are inspired by the two corresponding algorithms in \cite{FK96}, for \textsc{Monotone Boolean Duality Testing}, and can be thought of as generalizations. The first algorithm is simpler and exploits the idea of the existence of a large degree vertex in any non-colorable instance. 
By considering all possible admissible colorings of such  a vertex we can remove a large fraction of the edges and recurse on substantially smaller-size problems.
Unfortunately, the degree of the high-degree vertex is only large enough to guarantee a bound of $O(m^{\log^2m})$ (assuming $k$ and $c$ are fixed). The second algorithm is more complicated and considers both scenarios when there is a high-degree vertex and when there are none (where now the threshold for "high" is higher). If there is no high-degree vertex, then we can find a "balanced-set" which induces a constant number of edges. Then a decomposition can be obtained based on this set.     
 
\section{Solving \PKC\ in Quasi-polynomial Time} 
We give two lemmas that show the existence of a large degree vertex, unless the hypergraph is easily colorable.
\begin{lemma}\label{l1}
Let $\cH\subseteq 2^V$ be a given $c$-intersecting hypergraph, $\cL:V\to2^{[k]}$ be a mapping, and $\chi:V\to[0:k]$ be a proper partial $\cL$-coloring of $\cH$ such that $|\cH_0|> 2c$. 
Then either 
\begin{itemize}
\item[(i)] there is a vertex $v \in V_0$ with $\deg_{\cH_0}(v)> \frac{|\cH_0|}{2\log_{\nu}(m\kappa)}$, or 
\item[(ii)] an $\cL$-coloring $\chi_0: V_0\to[k]$, such that $\chi\cup\chi_0$ is a proper $\cL$-coloring of $\cH$, can be found in $O(\rho|V_0|m)$ time.
\end{itemize}
\end{lemma}
\proof
Let $H_{\min}$ be an edge in $\bigcup_{i=0}^k\cH_i^{V_0}$ of minimum size. Assume, without loss of generality, that $|H_{\min}|\ge2$.
Pick a random $\cL$-coloring $\chi_0:V_0\to[k]$ by assigning, independently for each $v\in V_0$, $\chi_0(v)=i\in L(v)$ with probability $\frac{1}{|\cL(v)|}$. 
Then, for an edge $H\in\cH_0$, $$\Pr[H\text{ is monochromatic}]=|\bigcap_{v\in H}L(v)|\cdot\prod_{v\in H}\frac{1}{|\cL(v)|}\le \kappa\cdot\left(\frac{1}{\nu}\right)^{|H|},$$ and for $H\in\cH_i$, $i\in [k]$, $$\Pr[H\text{ is monochromatic}]\le\prod_{v\in H\cap V_0}\frac{1}{|\cL(v)|}\le\left(\frac{1}{\nu}\right)^{|H\cap V_0|}.$$ It follows that
\begin{eqnarray*}
\EE[\text{\# monochromatic }H\in\cH]&=&\sum_{H\in\cH}\Pr[H\text{ is monochromatic}]\\
&\le&\kappa\sum_{H\in\cH_0}\left(\frac{1}{\nu}\right)^{|H|}+\sum_{i=1}^k\sum_{H\in\cH_i}\left(\frac{1}{\nu}\right)^{|H\cap V_0|}\le m\kappa\left(\frac{1}{\nu}\right)^{|H_{\min}|}.
\end{eqnarray*} 
Thus if $m\kappa\left(\frac{1}{\nu}\right)^{|H_{\min}|}<1$, then there is a proper $\cL$-coloring $\chi':=\chi\cup\chi_0$ of $\cH$, which can be found by the method of conditional expectations in time $O(\rho|V_0|m)$. Let us therefore assume for the rest of this proof that $|H_{\min}|\le \log_{\nu}(m\kappa)$.

Let $v_{\max}$ be a vertex maximizing $\deg_{\cH_0}(v)$ over $v\in H_{\min}$. Then \raf{intersecting} implies that  
\begin{align*}
|\cH_0|&\le\left|\bigcup_{v\in H_{\min}}\{H\in\cH_0:~v\in H\}\right|+c\le\sum_{v\in H_{\min}}|\{H\in\cH_0:~v\in H\}|+c\\&=\sum_{v\in H_{\min}}\deg_{\cH_0}(v)+c\le|H_{\min}|\deg_{\cH_0}(v_{\max})+c.
\end{align*}
Consequently, $\deg_{\cH_0}(v_{\max})\ge\frac{|\cH_0|-c}{|H_{\min}|}>\frac{|\cH_0|}{2\log_{\nu}(m\kappa)}.$ 
\qed

\begin{lemma}\label{l2}
	Let $\cH\subseteq 2^V$ be a given hypergraph $c$-intersecting hypergraph, % of minimum edge-size $2$, 
	$\cL:V\to2^{[k]}$ be a mapping, and $\chi:V\to[0:k]$ be a proper partial $\cL$-coloring of $\cH$ such that $|\cH_0|=0$, and for all $i\in[k]$, either $|\cH_i|=0$ or $|\cH_i|>2c$. 
	Then either 
	\begin{itemize}
	\item[(i)] there is a vertex $v \in V_0$ and $i,j\in[k]$, $j\ne i$, such that $\deg_{\cH_i}(v)> \frac{|\cH_i|}{2\log_\nu m}$ and $\deg_{\cH_j}(v)\ge 1$, 
	or \item[(ii)] an $\cL$-coloring $\chi_0: V_0\to[k]$, such that $\chi\cup\chi_0$ is a proper $\cL$-coloring of $\cH$, can be found in $O(\rho|V_0|m)$ time.
	\end{itemize}
\end{lemma}
\proof
Let  $H_{\min}$ be an edge in $\bigcup_{i=1}^k\cH_i^{V_0}$ of minimum size.
Note that \raf{intersecting} implies:
\begin{equation}\label{transversality}
%\text{(transversality)}~~
\forall H\in\cH_i:~H\cap H'\cap V_0\neq\emptyset~\text{ for all but at most $c$  edges }H'\in\bar{\cH}_i,
\end{equation}
since $\{i\}=\chi(H\setminus V_0)\ne\chi(H'\setminus V_0)=\{j\}$ for all $H\in\cH_i$ and $H'\in\cH_j$, for $i\ne j$. 

If there is an $i\in[k]$ such that $\cH_j=\emptyset$ for all $j\in[k]\setminus\{i\}$ then an $\cL$-coloring satisfying (ii) can be found by choosing arbitrarily $\chi(v)\in\cL(v)\setminus\{i\}$ for $v\in V_0$. Assume therefore that $\cH_i\ne\emptyset$ for at least two distinct indices $i\in[k]$.
Pick a random $\cL$-coloring $\chi_0:V_0\to[k]$ by assigning, independently for each $v\in V_0$, $\chi(v)=i\in\cL(v)$ with probability $\frac{1}{|\cL(v)|}$. Then
\begin{eqnarray*}
	\Pr[\exists i\in[k],~ H\in\cH_i:~\chi(H_i)=\{i\}]&\le&\sum_{i=1}^k\sum_{H\in\cH_i}\Pr[\chi(H)=\{i\}]\\
	&\le&
	\sum_{i=1}^k\sum_{H\in\cH_i}\prod_{v\in H\cap V_0}\frac{1}{|\cL(v)|}
	\le  m\left(\frac{1}{\nu}\right)^{|H_{\min}|}.
\end{eqnarray*} 
Thus if $m\left(\frac{1}{\nu}\right)^{|H_{\min}|}<1$, then there is an $\cL$-coloring satisfying (ii), which can be found by the method of conditional expectations in time $O(\rho |V_0|m)$. Let us therefore assume for the rest of this proof that $|H_{\min}|\le \log_\nu m$.

Let $j$ be such that $H_{\min}\in\cH_j^{V_0}$, and $v_{\max}$ be a vertex maximizing $\deg_{\bar\cH_j}(v)$ over $v\in H_{\min}$. Then \raf{transversality} implies that 
\begin{align*}
|\bar\cH_j|&=\left|\bigcup_{v\in H_{\min}}\{H\in\bar\cH_j:~v\in H\}\right|+c\le\sum_{v\in H_{\min}}|\{H\in\bar\cH_j:~v\in H\}|+c\\&=\sum_{v\in H_{\min}}\deg_{\bar\cH_j}(v)+c\le|H_{\min}|\deg_{\bar\cH_j}(v_{\max})+c.
\end{align*}
Consequently, $\sum_{i\ne j}\deg_{\cH_i}(v_{\max})=\deg_{\bar\cH_j}(v_{\max})\ge\frac{|\bar\cH_j|-c}{|H_{\min}|}\ge\frac{|\bar\cH_j|-c}{\log_\nu m}=\frac{\sum_{i\ne j}|\cH_i|-c}{\log_\nu m}\ge \frac{\sum_{i\ne j}(|\cH_i|-c)}{\log_\nu m}>\frac{\sum_{i\ne j}|\cH_i|}{2\log_\nu m},$ from which it follows that 
$\max_{i\ne j}\frac{\deg_{\cH_i}(v_{\max})}{|\cH_i|}\ge\frac{\sum_{i\ne j}\deg_{\cH_i}(v_{\max})}{\sum_{i\ne j}|\cH_i|}>\frac{1}{2\log_\nu m}$. 
\qed

\begin{algorithm}[!htb]
	\caption{ \PKC-A$(\cH,\cL,\chi)$} \label{PKC-alg-A}
\begin{algorithmic}[1]
\Require A $c$-intersecting hypergraph $\cH\subseteq 2^V$, a mapping $\cL:V\to2^{[k]}$, and a proper partial $\cL$-coloring $\chi:V\to[0:k]$
\Ensure A partial proper $\cL$-coloring $\chi:V\to[k]$ of $\cH$
\State $V_0:=V_0(\chi)$; $\cH:=\cH(\chi)$%; $\cH_i:=\cH_i(\chi)$ for $i\in[0:k]$  
\If{$|\cH_i|=0$ for all $i\in[0:k]$}
\State\textbf{stop} /* A proper $\cL$-coloring has been found */ 
\EndIf
\If{$|\cH_0|>\delta:=\max\{2c,\rho^{2}\}$} /* Phase I */
     \If {there is $v\in V_0$ satisfying condition (i) of Lemma~\ref{l1}}\label{s1-1}
   \For{each proper assignment $\chi_0(v)\in\cL(v)$} \label{s1-2}
   \State \textbf{call} \PKC-A$(\cH,\cL,\chi\cup\chi_0)$ \label{s1-3}
   \EndFor
   \Else
   \State Let $\chi_0:V_0\to[k]$ be a coloring computed as in (ii) of Lemma~\ref{l1} \label{s1-5}
   \State Set $\chi:=\chi\cup\chi_0$; \textbf{stop} /* A proper $\cL$-coloring has been found */ \label{s1-6}
   \EndIf 
\Else /* Phase II */
 \If{there is $i\in[0:k]$ such that $1\le|\cH_i|\le \delta$}  \label{s1-7}
\For{and each proper $i$-simple assignment $\chi_0$} /* Clean-up */\label{s1-7-}
\State \textbf{call} \PKC-A$(\cH,\cL,\chi\cup\chi_0)$ \label{s1-8}
\EndFor
\EndIf   
   \State Same as in steps \ref{s1-1}-\ref{s1-6} of Phase I (applying Lemma~\ref{l2} instead)\label{s1-9}
\EndIf    
   \State\Return  \label{s1-10}
\end{algorithmic}
\end{algorithm}

The algorithm for solving \PKC\ is given as Algorithm~\ref{PKC-alg-A}, which is called initially with $\chi\equiv 0$. The algorithm terminates either with a proper $\cL$-coloring of $\cH$, or with a partial $\cL$-coloring with some unassigned vertices, in which case we conclude that no proper $\cL$-coloring of $\cH$ exists.

The algorithm proceeds in at two phases. As long as the number of edges with no assigned colors is a above a certain threshold $\delta$, that is $|\cH_0|>\delta$, the algorithm is still in phase I; otherwise it proceeds to phase II.
In a general step of  of phase I (resp., phase II), the algorithm picks a vertex $v$ satisfying condition (i) of Lemma~\ref{l1} (resp., Lemma~\ref{l2}) and iterates over all feasible assignments of colors to $v$, that result in no monochromatic edges (line~\ref{s1-3}); if no such $v$ can be found, the algorithm concludes with a proper $\cL$-coloring. In each iteration, any edge that becomes non-monochromatic is removed and the algorithm recurses on the updated sets of hypergraphs.
If none of the recursive calls yields a feasible extension of the current proper partial $\cL$-coloring $\chi$, we unassign vertex $v$ and return (line~\ref{s1-10}). At the beginning of each recursive call in phase II, we preform a ''clean-up" step (lines~\ref{s1-7}-\ref{s1-8}) by trying all possible $i$-simple assignments for hypergrpahs $\cH_i$ with $|\cH_i|$ sufficiently small. This allows us to start phase II with $|\cH_0|=0$ and to keep only hypergrpahs $\cH_i$ whose size is above the threshold $\delta$.   

\medskip

To analyze the running time of the algorithm, let  
us measure the "volume" of a subproblem with input $(\cH,\cL,\chi)$, in phase I by $\mu_1=\mu_1(\cH,\chi):=|\cH_0(\chi)|$,
and in phase II by 
\begin{equation}
\label{measure}
\mu_2=\mu_2(\cH,\chi):=\prod_{i=0}^k\max\{|\cH_i(\chi)|,1\}.
\end{equation} 
The recursion stops when $\mu_1(\cH,\chi)=\mu_2(\cH,\chi)=0$ (meaning that $|\cH_i|=0$ for all $i\in[0:k]$, and hence, the algorithm managed to completely color all vertices), an $\cL$-coloring satisfying condition (i) of Lemmas~\ref{l1} or \ref{l2} is found (in lines~\ref{s1-6} or \ref{s1-9}), or when no proper extension of the current partial coloring $\cX$ can be found (no proper assignment exists in lines~\ref{s1-2}, \ref{s1-7-}, or~\ref{s1-9}). 

\begin{lemma}
Algorithm~\ref{PKC-alg-A} solves problem \PKC\ in time $(\rho n)^{O((\rho^2+c)k)}(\kappa m)^{h}$, where $h:=O(\frac{\log^2(\rho+c)+k\log(\rho+c)\log m+k^2\log^2 m}{\log\nu})$.
\end{lemma}
\proof
Let $\epsilon:=\frac{1}{2\log_\nu(\kappa m)}$, $\alpha=\frac{1}{1-\epsilon}$  and $\delta:=\max\{2c,\rho^{2}\}$. We may assume that $\epsilon\le \frac12$, since otherwise $m\le\rho$, implying that the algorithm would terminate in $O((\rho n)^{2\rho})$ time after trying all simple $0$-assignments in lines~\ref{s1-1}-\ref{s1-2}.  Note that this implies that $\delta\ge\alpha^2$ as $\rho\ge 2\ge\frac{1}{1-\epsilon}=\alpha$.

Consider the recursion tree $\bT$ of the algorithm. Let $\bT_1$ (resp., $\bT_2$) be the subtree (resp., sub-forest) of $\bT$ belonging to phase I (resp., phase II) of the algorithm. Note that $\bT_2$ consists of maximal sub-trees of $\bT$, each of which is rooted at a leaf in $\bT_1$.  For $\mu_1\ge 0$ (resp., $\mu_2\ge 0$ and $t\in[0:k]$), let use denote by $A_1(\mu_1)$ (resp., $A_2(\mu_2,t)$) be the total number of nodes in $\bT_1$ (resp., $\bT_2$) that result from a subproblem of volume $\mu_1$ (resp., $\mu_2$ with $|\{i\in[0:k]:~|\cH_i(\chi)|\ge1\}|=t$) in phase I (resp., phase II). For each recursive call of the algorithm, we obtain a recurrence on $A_1(\mu_1)$ (resp., $A_2(\mu_2,t)$), as explained in the following. Naturally, we assume that $A_1(\mu_1)$ (resp., $A_2(\mu_2,t)$) is monotonically increasing in $\mu_1$(resp., in both $\mu_2$ and $t$). For simplicity and to avoid confusion, we denote by $\cH_i,\mu_1,\mu_2,t$ and $\cH_i',\mu'_1,\mu_2',t'$ the hypergraphs, volumes and the number of non-empty hypergraphs, in the current and next recursive calls, respectively. For the sake of the analysis, without loss of generality, we assume throughout that the algorithm does not terminate on a ``forced stop" as in line~\ref{s1-6}.
\begin{claim}\label{cl1}
	$A_1(\mu_1)\le\mu_1^{\log_{\alpha} \rho}$.
\end{claim}
\proof
Let $v\in V$ be the vertex chosen in line~\ref{s1-1}. Since $v$ is a large-degree vertex (with respect to $\cH_0$) which receives a color, $|\cH_0'|\le\big(1-\frac{1}{2\log_{\nu}(m\kappa)}\big)|\cH_0|$.
Thus, for a non-leaf node of $\bT_1$, we have the recurrence:
\begin{equation}\label{rA-1}
A_1(\mu_1)\le \rho\cdot A_1((1-\epsilon)\mu_1)+1.
\end{equation}
At leaves we have $\mu_1\le\delta$. It follows that the depth $d(\mu_1)$ of the recursion subtree of a node (in $\bT_1$) of volume $\mu_1$ is at most $\log_{\alpha} \frac{\mu_1}{\delta}+1$, where $\mu_1=|\cH_0|$ is the initial volume, and hence the total number $A_1(\mu_1)$ of nodes is bounded by $\frac{\rho^{d(\mu_1)+1}-1}{\rho-1}\le \rho^2\Big(\frac{\mu_1}{\delta}\Big)^{\log_{\alpha} \rho}\le\mu_1^{\log_{\alpha} \rho}$ (as $\rho\ge\alpha$ and $\delta\ge \rho^2$). 
\qed

\begin{claim}\label{cl2}
	$A_2(\mu_2,t)\le(\rho n)^{2\delta\cdot(t+1)}\mu_2^{\log_{\alpha} \mu_2}$.
\end{claim}
\proof
There are two possible locations in which a recursive call can be initiated in phase II: 
\medskip

\noindent{\it Line~\ref{s1-8}}: Since $t'\le t-1$, as we remove at least one hypergraph $\cH_i$ by trying all $i$-simple assignments whose number is at most $(|V_0|\rho)^{2|\cH_i|}$, where $|\cH_i|\le \delta$, we get the recurrence   
\begin{equation}\label{rA-2}
A_2(\mu_2,t)\le(\rho n)^{2\delta}A_{2}(\mu_2,t-1)+1.
\end{equation} 

\medskip

\noindent{\it Line~\ref{s1-9}} (the part corresponding to line~\ref{s1-3}): Let $v\in V$ be the vertex chosen before the recursive call (as in line~\ref{s1-1}), that is, 
$v$ satisfies condition  (i) of Lemma~\ref{l2}, and let $i,j\in[k]$ be such that $i\ne j$, $\deg_{\cH_i}(v)\ge \frac{1}{2\log_\nu m}|\cH_i|$ and $\deg_{\cH_j}(v)\ge 1$. There are $|\cL(v)|$ recursive calls that will be initiated from this point, corresponding to $\ell\in \cL(v)$; consider the $\ell$th recursive call. If $\ell\ne i$ then setting $\chi(v)=\ell$ will result in deleting all the edges containing $v$ from $\cH_i$. Thus, $\mu'_2\le(1-\frac{1}{2\log_\nu m})\mu_2$ if $|\cH_i'|>0$ and $\mu_2'\le\frac{\mu_2}{\delta}$, $t'\leq t-1$ if $|\cH_i'|=0$. In both cases, we get $\mu_2'\le(1-\epsilon)\mu_2$ (as $\frac1{\delta}<\frac12\le1-\epsilon$). On the other hand, if $\ell=i$, then at least one edge in $\cH_{j}$ will be deleted, yielding $\mu'_2\le \mu_2-1$, or $\mu_2'\le\frac{\mu_2}{\delta}$ and $t'\le t-1$, depending on whether $|\cH_j'|>0$ or $|\cH_j'|=0$. Again in both cases, for $\mu_2>\delta$, wet get $\mu_2'\le\mu_2-1$ (as $\frac{1}{\delta}<1-\frac{1}{\delta}<1-\frac{1}{\mu_2}$). Consequently, for $\mu_2>\delta$, we get the recurrence:
\begin{equation}\label{rA-3}
A_2(\mu_2,t)\le(\rho-1)\cdot A_{2}((1-\epsilon)\mu_2,t)+A_{2}(\mu_2-1,t)+1.
\end{equation}
By definition,  $A_{2}(\mu_2,0)=1$, for $\mu_2\ge0$. We will prove by induction on $t=1,\ldots,k$ and $\mu_2\ge1$ that 
\begin{equation}\label{rec-solve-A}
A_2(\mu,t) \le P_{t+1}\mu_2^{\log_\alpha \mu_2}, 
\end{equation}where $P_{t+1}:=\frac{R^{t+1}-1}{R-1}$ and $R:=(\rho n)^{2\delta}$. We consider 2 cases:

\smallskip

\noindent{Case 1.}~ $1\le\mu_2\le\delta$: Then $|\cH_i(\chi)|\le\delta$ for all $i\in[0:k]$, and recurrence~\raf{rA-2} applies iteratively until we get $t=0$.  By the recurrence, $A_2(\mu_2,1)\le R+1\le P_{2}\mu^{\log_\alpha \mu}$, giving the base case ($t=1$), and by induction on $t$, 
\begin{align*}
A_2(\mu_2,t)&\le R\left(P_t\mu_2^{\log_\alpha \mu_2}\right)+1\le \mu_2^{\log_\alpha \mu_2}\left( RP_t+1\right)= P_{t+1}\mu_2^{\log_\alpha \mu_2}.
\end{align*}

\medskip

\noindent{Case 2.}~ $\mu_2>\delta$:  If the recurrence in~\raf{rA-2} applies then the same induction proof (on $t$) in case 1 gives the required bound.
Consider, thus, the recurrence in~\raf{rA-3} and apply induction on $\mu_2$:
\begin{eqnarray*}
	A_2(\mu_2,t)&\le& (\rho-1)   P_{t+1}\left((1-\epsilon)\mu_2\right)^{\log_{\alpha}((1-\epsilon)\mu_2)}
	+P_{t+1}(\mu_2-1)^{\log_{\alpha} (\mu_2-1)}+1\\
	&\le& P_{t+1}\frac{\rho-1}{(1-\epsilon)\mu_2}\cdot\frac{1}{\mu_2}\cdot\mu_2^{\log_\alpha \mu_2}+P_{t+1}(\mu_2-1)^{\log_{\alpha}\mu_2}+1\\
	&<& P_{t+1}\mu_2^{\log_{\alpha} \mu_2}\left(\frac{1}{\mu_2}+\left(1-\frac{1}{\mu_2}\right)^{\log_\alpha\mu_2}+\frac{1}{P_{t+1}\mu_2^{\log_{\alpha} \mu_2}}\right)~~\text{ ($\because\mu_2\ge\delta$ and  $\epsilon\le\frac12$)}.\\
	&\le&P_{t+1}\mu_2^{\log_{\alpha} \mu_2}\left(\frac{1}{\mu_2}+\left(1-\frac{1}{\mu_2}\right)^{2}+\frac{1}{\mu_2^2}\right) ~~\text{ ($\because\mu_2\ge\delta\ge\alpha^2$ and hence $\log_{\alpha}\mu_2\ge 2$  )}\\
	&\le&P_{t+1}\mu_2^{\log_{\alpha} \mu_2} ~~\text{ ($\because\mu_2\ge\delta>2$)}.
\end{eqnarray*}
\qed

\medskip

Using the bounds $\mu_1\le m$, and  
$$\mu_2\le \prod_{i=0}^k|\cH_i|\le \delta\cdot\left(\frac{\sum_{i=1}^k|\cH_i|}{k}\right)^{k}\le\delta\cdot\left(\frac{m}{k}\right)^{k},
$$
we get $A_1(\mu_1)=m^{O(\log \rho\cdot\log_{\nu}(m\kappa))}$ and $A_2(\mu_2,k+1)=(\rho n)^{2\delta\cdot(k+1)}\left(\delta \left(\frac{m}{k}\right)^k\right)^{ O\big(\log\big(\delta \left(\frac{m}{k}\right)^k\big)\cdot\log_{\nu}(m\kappa)\big)}$. Putting Claims~\ref{cl1} and~\ref{cl2} together, and noting that at internal nodes the running time is $O(\rho nm)$, and that the roots of the maximal sub-trees in $\bT_2$ are the leaves of $\bT_1$, the lemma follows.
\qed
 
\section{A More Efficient Algorithm}
%\subsection{Preliminaries}

%Philosophy behing the second algorithm
When $k=O(1)$ and $c=O(1)$, the algorithm presented in the previous section for \PKC\ has running time $n^{O(1)}m^{O(\log^2m)}$. Moreover, the recursion tree can have depth $\Omega(m)$. In this section, we give an algorithm with running time $(nm)^{o(\log m)}$ and recursion-tree depth $\polylog(m,n)$ (for $k=O(1)$ and $c=\polylog(m,n)$), thus proving Theorem~\ref{t1}. The speedup comes from the fact that the algorithm may assign one color to a complete set of vertices in one time step, rather than to a single vertex as in the previous algorithm. In fact, as we shall see below, the algorithm may ``probe"  a color assignment on a certain set (the set $V_0\setminus S$ in Lemmas~\ref{l3} and ~\ref{l4}); if such an assignment cannot be completed to a proper coloring for the whole hypergraph, the information gained from such a ``failure" turns out to be useful for restricting the set of color assignments the algorithm should try next. In general, such a probing strategy may be expensive, but as we shall see below, we can use a set $S$ satisfying some  ``balancing" condition (see Lemmas~\ref{l5} and~\ref{l6}) to ensure that the increase in the running time from probing is offset by the amount of information gained. 
 
 \medskip
 
For a hypergraph $\cH\subseteq 2^V$ and a positive number $\epsilon\in(0,1)$, denote by $T(\cH,\epsilon)$ the subset $\{v\in V:~\deg_{\cH}(v)> \epsilon|\cH|\}$ of "high" degree vertices in $\cH$.
Given $\epsilon',\epsilon''\in(0,1)$, $\epsilon'<\epsilon''$, let us call an {\it $(\epsilon',\epsilon'')$-balanced set} with respect to $\cH$, any set $S\subseteq V$ such that $\epsilon'|\cH|\le|\cH_S|\le \epsilon''|\cH|$.  

\begin{proposition}[\cite{E08}]\label{p2}
Let $\epsilon_1,\epsilon_2\in(0,1)$ be two given numbers such that, $\epsilon_1<\epsilon_2$ and $T=T(\cH,\epsilon_1)$ satisfies $|\cH_{T}|\leq(1-\epsilon_2)|\cH|$. Then there exists a $(1-\epsilon_2,1-(\epsilon_2-\epsilon_1))$-balanced set $S\supseteq T$ with respect to $\cH$. Such a set $S$ can be found in $O(nm)$ time.
\end{proposition}
\proof
Let $\{v_1,\ldots,v_l\}$ be 
an arbitrary order of the vertices of $\bar T$ and find the index $j\in[l-1]$, 
such that 
\begin{equation}\label{part}
|\cH_{V\smallsetminus\{v_1,\ldots,v_j\}}|>(1-\epsilon_2) |\cH|\mbox{ and }
|\cH_{V\smallsetminus\{v_1,\ldots,v_{j+1}\}}|\le(1-\epsilon_2) |\cH|.
\end{equation}
The existence of such $j$ is guaranteed by the facts that $\deg_{\cH}(v_1)\le\epsilon_1|\cH|< \epsilon_2|\cH|\leq|\cH(\bar T)|$. 
Finally, we let $S=V\smallsetminus\{v_1,\ldots,v_j\}$. Since $\deg_{\cH}(v_{j+1})\le\epsilon_1|\cH|$, it follows from \raf{part} that $|\cH_{S}|<(\epsilon_1+1-\epsilon_2)|\cH|$, implying that $S$ is indeed a balanced superset of $T$. 
\qed

\begin{lemma}\label{l3}
Let $\cH\subseteq 2^V$ be a $c$-intersecting hypergraph, $\cL:V\to2^{[k]}$ be a mapping, $\chi:V\to[0:k]$ be a proper partial $\cL$-coloring of $\cH$, and $S\subseteq V_0$ be a given set of vertices such that $\cH_{V\setminus S}=\emptyset$ (equivalently, $\emptyset\not\in\cH^S$). Fix an arbitrary (proper) coloring $\chi_p:V_0\setminus S\to[k]$ and let $\widehat \chi:=\chi\cup\chi_p$. Then, $\chi$ is extendable to a proper $\cL$-coloring of $\cH$ if and only if either
\begin{itemize}
\item[(i)] $\chi$ is extendable to a proper $\cL$-coloring $\chi'$ for $\cH$, with $\chi'(V_0\setminus S)=\chi_p$, or
\item[(ii)] $\exists H\in\cH\setminus(\cH_0)_S: |\widehat\chi(H\setminus S)|=1$, $\chi$ is extendable to a proper $\cL$-coloring $\chi'$ for $\cH$ such that $\chi'(H\cap S)=\widehat\chi(H\setminus S)$.
\end{itemize} 
\end{lemma}
\proof
First note that $\chi_p$ does not introduce any monochromatic edges as $\cH_{V\setminus S}=\emptyset$.
Suppose that $\chi$ is extendable to a proper $\cL$-coloring $\chi'$ for $\cH$. The fact that (i) is not satisfied means that there is an $H\in\cH\setminus(\cH_0)_S$, such that (in any proper extension $\chi'$ of $\chi$), $\chi'$ assigns a single color to all the vertices in $H\cap S$, which is exactly the color assigned by $\chi\cup\chi_p$ to all vertices in $H\setminus S$, and hence (ii) is satisfied. 

Conversely, if (i) or (ii) hold, then trivially, there is an $\cL$-coloring extension $\chi'$ of $\chi$ that properly colors $\cH$.  
\qed

\begin{lemma}\label{l4}
	Let $\cH\subseteq 2^V$ be a $c$-intersecting hypergraph, $\cL:V\to2^{[k]}$ be a mapping, $\chi:V\to[0:k]$ be a proper partial $\cL$-coloring of $\cH$, and $S\subseteq V_0$ be a given set of vertices such that, for some $i\in[k]$, $\emptyset\not\in\bar\cH_i^{S}$ (equivalently, $(\cH_j)_{V\setminus S}=\emptyset$ for all $j\ne i$). Fix $\chi_p:V_0\setminus S\to[k]$ by setting $\chi_p(v)\in\cL(v)\setminus\{i\}$ arbitrarily for $v\in V_0\setminus S$, and let $\widehat \chi:=\chi\cup\chi_p$. Then $\chi$ is extendable to a proper $\cL$-coloring of $\cH$ if and only if either
\begin{itemize}\item[(i)] $\chi$ is extendable to a proper $\cL$-coloring $\chi'$ for $\cH$, with $\chi'(V_0\setminus S)=\chi_p$, or
\item[(ii)] $\exists H\in\cH\setminus\cH_i: |\widehat\chi(H\setminus S)|=1$, $\chi$ is extendable to a proper $\cL$-coloring $\chi'$ for $\cH$ such that $\chi'(H\cap S)=\widehat\chi(H\setminus S)$. 
\end{itemize} 
\end{lemma}
\proof
First note that $\chi_p$ does not introduce any monochromatic edges as $H\cap S\ne\emptyset$ for all $H\in\cH\setminus\cH_i$.
Suppose that $\chi$ is extendable to a proper $\cL$-coloring $\chi'$ for $\cH$. If (i) is not satisfied then there is an $H\in\cH_j$ for some $j\ne i$, such that $\chi'$ assigns a single color to all the vertices in $H\cap S$, which is exactly the color assigned by $\chi\cup\chi_p$ to all vertices in $H\setminus S$, and hence (ii) is satisfied. 

Conversely, if (i) or (ii) hold, then trivially, there is an $\cL$-coloring extension $\chi'$ of $\chi$ that properly colors $\cH$.  
\qed

\begin{lemma}\label{l5}
	Let $\cH\subseteq 2^V$ be a $c$-intersecting hypergraph, $\cL:V\to2^{[k]}$ be a mapping, $\chi:V\to[0:k]$ be a proper partial $\cL$-coloring of $\cH$ such that $|\cH_0|>0$, $T(\cH_0,\epsilon_1)=\emptyset$, and $\epsilon_1,\epsilon_2\in(0,1)$ be two given numbers satisfying $\epsilon_1<\frac{\epsilon_2}{1+c}$. Then there is a $(1-\epsilon_2,1-(\epsilon_2-(1+c)\epsilon_1))$-balanced set $S\subseteq V_0$ with respect to $\cH_0$ such that $\cH_{V\setminus S}=\emptyset$. 
\end{lemma}
\proof
We start with a $(1-\epsilon_2,1-(\epsilon_2-\epsilon_1))$-balanced set $S'\subseteq V_0$, guaranteed by Proposition~\ref{p2}. Since $(\cH_0)_{S'}\ne\emptyset$, by \raf{intersecting} we have $|\cH_{V\setminus S'}|\le c$. Let $S$ be the set obtained by appending to $S'$ a single vertex from each edge (if any) in $\cH_{V\setminus S'}^{V_0\setminus S}$ (that is, vertices are added from $V_0\setminus S'$). Then, since the degree of each appended vertex is no more than $\epsilon_1|\cH_0|$ (as $S'\supseteq T(\cH_0,\epsilon_1)$), we get 
$$(1-\epsilon_2)|\cH_0|\le|(\cH_0)_{S'}|\le|(\cH_0)_{S}|\le |(\cH_0)_{S'}|+c\cdot\epsilon_1|\cH_0|\le\big(1-(\epsilon_2-\epsilon_1)+c\cdot\epsilon_1\big)|\cH_0|.$$
\qed

\begin{lemma}\label{l6}
Let $\cH\subseteq 2^V$ be a $c$-intersecting hypergraph, $\cL:V\to2^{[k]}$ be a mapping, $\chi:V\to[0:k]$ be a proper partial $\cL$-coloring of $\cH$ such that $|\cH_i|>2c$ for at least two $i$'s, and $\epsilon_1,\epsilon_2\in(0,\frac12]$ be two given numbers satisfying $\epsilon_1<\epsilon_2$. Then either (i) there is $v\in V_0$ and $i\ne j$ such that $\deg_{\cH_i}(v)\ge\epsilon_1|\cH_i|$ and $\deg_{\cH_j}(v)\ge\epsilon_1|\cH_j|$, or (ii) there is a $(1-\epsilon_2,1-(\epsilon_2-(1+c)\epsilon_1))$-balanced set $S\subseteq V_0$ with respect to $\cH_i^{V_0}$ for some  $i\in[k]$, such that $(\cH_j)_{V\setminus S}=\emptyset$ for all $j\ne i$.  
\end{lemma}
\proof
For any $i\ne j$ such that $|\cH_i|>2c$ and $|\cH_j|>2c$, let $T_i:=T(\cH_i^{V_0},\epsilon_1)$ and $T_j:=T(\cH_j^{V_0},\epsilon_1)$. If $T_i\cap T_j\ne\emptyset$ then any $v$ in this intersection will satisfy (i). Otherwise, \raf{intersecting} implies that either $|(\cH_i^{V_0})_{T_i}|\le c\le(1-\epsilon_2)|\cH_i^{V_0}|$ or $|(\cH_j^{V_0})_{T_j}|\le c\le(1-\epsilon_2)|\cH_j^{V_0}|$ (as any $H\in\cH_i$ and $H'\in\cH_j$ cannot intersect outside $V_0$), in which case a $(1-\epsilon_2,1-(\epsilon_2-\epsilon_1))$-balanced set $S'$ with respect to $\cH_i^{V_0}$ or $\cH_j^{V_0}$, respectively, can be obtained by Proposition~\ref{p2}. Suppose $S'$ was obtained w.r.t. $\cH_i^{V_0}$ but there are some edges in $\bar \cH_i^{V_0}$ that are induced by $V_0\setminus S'$. Let $S$ be the set obtained by appending to $S'$ a single vertex from each such edge (that is, vertices are added from $V_0\setminus S'$). Then, since the number of such edges is at most $c$ (by \raf{intersecting}) and the degree of each appended vertex is no more than $\epsilon_1|\cH_i^{V_0}|$ (as $S'\supseteq T(\cH_i^{V_0},\epsilon_1)$), we get 
$$(1-\epsilon_2)|\cH_i^{V_0}|\le|(\cH_i^{V_0})_{S'}|\le|(\cH_i^{V_0})_{S}|\le |(\cH_i^{V_0})_{S'}|+c\cdot\epsilon_1|\cH_i^{V_0}|\le\big(1-(\epsilon_2-\epsilon_1)+c\cdot\epsilon_1\big)|\cH_i^{V_0}|.$$
\qed

\begin{algorithm}[!htb]
	\caption{ \PKC-B$(\cH,\cL,\chi)$} \label{PKC-alg-B}
	\small
	\begin{algorithmic}[1]
		\Require A $c$-intersecting hypergraph $\cH\subseteq 2^V$, a mapping $\cL:V\to2^{[k]}$, and a proper partial $\cL$-coloring $\chi:V\to[0:k]$
		\Ensure A partial proper $\cL$-coloring $\chi:V\to[k]$ of $\cH$
    	\State $V_0:=V_0(\chi)$; $\cH:=\cH(\chi)$; $\mu_1:=\mu_1(\cH,\chi)$; $\mu_2:=\mu_2(\cH,\chi)$
    	\If{$|\cH_i|=0$ for all $i\in[0:k]$}
        	\State\textbf{stop} /* A proper $\cL$-coloring has been found */ 
    	\EndIf
		\If{$|\cH_0|>\delta(m)$} \label{s2-0} /* Phase I */ 
	      	\If {there is $v\in V_0$ such that      $\deg_{\cH_0}(v)\ge\epsilon_1(\mu_1)|\cH_0|$}\label{s2-1}
	         	\For{each proper assignment $\chi_0(v)\in\cL(v)$} \label{s2-2}
	              	\State \textbf{call} \PKC-B$(\cH,\cL,\chi\cup\chi_0)$ \label{s2-3}
	        	\EndFor
	    	\Else
		       \State Let $S$ be a $(1-\epsilon_2(\mu_1),1-(\epsilon_2(\mu_1)-(1+c)\epsilon_1(\mu_1)))$-balanced set computed as in Lemma~\ref{l5} w.r.t. $\cH_0$ \label{s2-5}
		       \For{each $v\in V_0\setminus S$} set $\chi_p(v)\in\cL(v)$ arbtitrarily \label{s2-5-}
		       \EndFor
	           \State \textbf{call} {\PKC-B$(\cH,\cL,\chi\cup\chi_p)$} \label{s2-6} /* Probe */
     	          \For{each $H\in\cH\setminus(\cH_0)_S$ such that $|\chi\cup\chi_p(H\setminus S)|=1$, and proper assignment $\chi_0(H\cap S):=\chi\cup\chi_p(H\setminus S)$}\label{s2-7}
     	          \State \textbf{call} \PKC-B$(\cH,\cL,\chi\cup\chi_0)$ \label{s2-8}
     	           \EndFor 		    
     	    \EndIf               
    \Else /* Phase II */
           \If{there is $i\in[0:k]$ such that $1\le|\cH_i|\le \delta(m^k)$}   \label{s2-10}
              \For{and each proper $i$-simple assignment $\chi_0$} /* Clean-up */
                   \State \textbf{call} \PKC-B$(\cH,\cL,\chi\cup\chi_0)$ \label{s2-11}
              \EndFor
          \EndIf   
               \If {there is $v\in V_0$ and $i\ne j$ such that $\deg_{\cH_i}(v)\ge\epsilon_1(\mu_2)|\cH_i|$ and $\deg_{\cH_j}(v)\ge\epsilon_1|\cH_j|$}\label{s2-12}
              \State Same as in steps \ref{s2-2}-\ref{s2-3} of Phase I\label{s2-13}
            \Else   
              \State Let $S$ be a $(1-\epsilon_2(\mu_2),1-(\epsilon_2(\mu_2)-(1+c)\epsilon_1(\mu_2)))$-balanced set computed as in Lemma~\ref{l6} w.r.t. $\cH_i^{V_0}$ for some $i\in[k]$\label{s2-14}
             \For{each $v\in V_0\setminus S$} set $\chi_p(v)\in\cL(v)\setminus\{i\}$ arbtitrarily \label{s2-15} 
             \EndFor
            \State \textbf{call} {\PKC-B$(\cH,\cL,\chi\cup\chi_p)$}  \label{s2-16} /* Probe */
               \For{each $j\ne i$, $H\in\cH_j$ and proper assignment $\chi_0(H\cap S):=\{j\}$}\label{s2-18}
                   \State \textbf{call} \PKC-B$(\cH,\cL,\chi\cup\chi_0)$\label{s2-19}
              \EndFor
		\EndIf 
		\EndIf
	 \State\Return\label{s2-20}
	\end{algorithmic}
\end{algorithm}

Again, the algorithm proceeds in two phases. As long as there is still a good number of edges with no assigned colors, the algorithm is still in phase I; otherwise it proceeds to phase II.
In a general step of phase I (resp., phase II), the algorithm tries, in line~\ref{s2-1} (resp., line~\ref{s2-12}), to find a vertex $v$ of large degree in $\cH_0$ (resp., in $\cH_i$ and $\cH_j$ for some $i\ne j$) and iterates over all feasible assignments of colors to $v$, that result in no monochromatic edges; if no such $v$ can be found then Lemma~\ref{l5} (resp.,  Lemma~\ref{l6}) guarantees the existence of a $(1-\epsilon_2,1-(\epsilon_2-(1+c)\epsilon_1))$-balanced set with respect to $\cH_0$ (resp., with respect to some $\cH_i^{V_0}$, $i\in[k]$), which is found in  line~\ref{s2-5} (resp, \ref{s2-14}). Lemma~\ref{l3} (resp., Lemma~\ref{l4}) then reduces the problem in the latter case to checking conditions (i) and (ii) of the lemma, which is done in lines \ref{s2-5-}-\ref{s2-8} (resp., \ref{s2-15}-\ref{s2-19}). If none of the recursive calls yields a feasible extension of the current proper partial $\cL$-coloring $\chi$, we unassign all the vertices colored in this call and return (line~\ref{s2-20}).  At the beginning of each recursive call in phase II, we preform a ''clean-up" step (lines~\ref{s2-10}-\ref{s2-11}) by trying all possible $i$-simple assignments for hypergrpahs $\cH_i$ with $|\cH_i|$ sufficiently small. This allows us to start phase II with $|\cH_0|=0$ and to keep only hypergrpahs $\cH_i$ whose size is above the prescribed threshold $\delta$.  
 
\medskip
As in the previous section, to analyze the running time of the algorithm, we  measure the volume of a subproblem in phase I by $\mu_1=\mu_1(\cH,\chi)=|\cH_0(\chi)|$, and in phase II by $\mu_2=\mu_2(\cH,\chi)$ given by \raf{measure}. 
The recursion stops when $|\cH_i(\chi)|=0$ for all $i\in[0:k]$, or no proper extension of the current partial coloring $\cX$ can be found. 

Given a subproblem of volume $\mu$, let $\epsilon(\mu):=\frac{4(c+1)\ln (2\rho)}{\xi(\mu)}$, where $\xi(\mu)$ is the unique positive root of the equation:
\begin{equation}\label{chi}
\left(\frac{1}{\epsilon(\mu)}\right)^{\xi(\mu)}=2\mu.
\end{equation} 
Note that $\xi(\mu)>4(c+1)\ln (2\rho)>1$, and hence $\epsilon(\mu)<1$, for $\mu\ge1$, and that (for constant $\rho$ and $c$) $\xi(\mu)\approx O\left(\frac{\log\mu}{\log\log\mu}\right)$.
We use in the algorithm: 
\begin{equation}\label{parameters}
\delta(\mu):=\frac{2(c+1)}{\epsilon(\mu)},~~\epsilon_1(\mu):=\frac{\epsilon(\mu)}{4(c+1)},~\text{ and }\epsilon_2(\mu):=\frac{\epsilon(\mu)}{2}.
\end{equation}

\begin{lemma}
	Algorithm~\ref{PKC-alg-B} solves problem \PKC\ in time $(\rho mn)^{h}$ where $h:=O(\frac{k^2\log_Hm}{\log_H\log_Hm})$ and $H:=4(c+1)\ln(2\rho)$.
\end{lemma}
\proof
Consider the recursion tree $\bT$ of the algorithm. Let $\bT_1$ (resp., $\bT_2$) be the subtree (resp., sub-forest) of $\bT$ belonging to phase I (resp., phase II) of the algorithm. For $\mu_1\ge 0$ (resp., $\mu_2\ge 0$ and $t\in[0:k]$), let use denote by $B_1(\mu_1)$ (resp., $B_2(\mu_2,t)$) be the total number of nodes in $\bT_1$ (resp., $\bT_2$) that result from a subproblem of volume $\mu_1$ (resp., $\mu_2$ with $|\{i\in[0:k]:~|\cH_i(\chi)|\ge1\}|=t$) in phase I (resp., phase II). For each recursive call of the algorithm, we obtain a recurrence on $B_1(\mu_1)$ (resp., $B_2(\mu_2,t)$), as explained in the following. Again, we assume that $B_1(\mu_1)$ (resp., $B_2(\mu_2,t)$) is monotonically increasing in $\mu_1$(resp., in both $\mu_2$ and $t$). Also, as before, we denote by $\cH_i,\mu_1,\mu_2,t$ and $\cH_i',\mu'_1,\mu_2',t'$ the hypergraphs, volumes and the number of non-empty hypergraphs, in the current and next recursive calls, respectively. 

\begin{claim}\label{cl3}
	 $B_1(\mu_1)\le\mu_1^{\xi(\mu_1)}$.
\end{claim}
\proof  
There are two possible locations in which a recursive call can be initiated in phase I: 
\medskip

\noindent{\it Line~\ref{s2-3}}: In this case, there is a vertex $v\in V_0$ such that $\deg_{\cH_0}(v)\ge\epsilon_1(\mu_1)|\cH_0|$, and we get $|\cH_0'|\le(1-\epsilon_1(\mu_1))$, and consequently the recurrence:
\begin{eqnarray}\label{recb1}
B_1(\mu_1)&\le& \rho\cdot B_1((1-\epsilon_1(\mu_1))\mu_1)+1\nonumber\\
&=&\rho\cdot B_1\Big(\Big(1-\frac{\epsilon(\mu_1)}{4(c+1)}\Big)\mu_1\Big)+1,  
\end{eqnarray}
since the recursion in line~\ref{s2-3} will exclude all the edges containing $v$ from $\cH_0'$. 

\medskip

\noindent{\it Lines~\ref{s2-6} and~\ref{s2-8}}: In this case, no large-degree vertex can be found. Then Lemma~\ref{l5} implies that there is a $(1-\epsilon_2(\mu_1),1-(\epsilon_2(\mu_1)-(1+c)\epsilon_1(\mu_1)))$-balanced set $S$, with respect to $\cH_0$, which is found in line~\ref{s2-5}. Then we apply Lemma~\ref{l3} which reduces the problem to one recursive call on the hypergraph $\cH$ in line~\ref{s2-6}, but after fixing the colors of all vertices in $V_0\setminus S$, and at most $|\cH\setminus(\cH_0)_S|$ recursive calls (in lines~\ref{s2-7}-\ref{s2-8}) on the hypergraphs obtained by fixing the color of one set $H\cap S$, for some $H\in\cH\setminus(\cH_0)_S$. Note that $S$ satisfies: $(1-\epsilon_2(\mu_1))|\cH_0|\le|(\cH_0)_{S}|\le(1-(\epsilon_2(\mu_1)-(1+c)\epsilon_1(\mu_1)))|\cH_0|$. In particular, there is are at least $(\epsilon_2(\mu_1)-(1+c)\epsilon_1(\mu_1))|\cH_0|$ edges $H\in\cH_0$ such that $H\setminus S\ne\emptyset$, and hence all of these edges will be removed from $\cH_0'$ in line~\ref{s2-6} giving $\mu_1'\le\mu_1\big(1-(\epsilon_2(\mu_1)-(1+c)\epsilon_1(\mu_1))\big)$. Moreover, in line~\ref{s2-8}, we will have $|\cH_0'|\leq|\cH_0\setminus(\cH_0)_S|+c$, as (by \raf{intersecting}) all but at most $c$ edges in $(\cH_0)_S$ have non-empty intersections with the set $H\cap S$, in the current iteration of the loop in line~\ref{s2-7}, all vertices of which are assigned the color $\chi\cup\chi_p(H\setminus S)$. Since $|\cH_0\setminus(\cH_0)_S|\le\epsilon_2(\mu_1)|\cH_0|$, we get the following recurrence for $|\cH_0|>\delta(m)\ge\delta(\mu_1)$:
\begin{eqnarray}\label{recb2}
B_1(\mu_1)&\le& B_1((1-(\epsilon_2(\mu_1)-(1+c)\epsilon_1(\mu_1))\mu_1)+\epsilon_2(\mu_1)\mu_1\cdot B_1\left(\Big(\epsilon_2(\mu_1)+\frac{c}{\delta(\mu_1)}\Big)\mu_1\right)+1\nonumber\\
&\le& B_1\Big(\Big(1-\frac{\epsilon(\mu_1)}{4}\Big)\mu_1\Big)+\frac{\epsilon(\mu_1)}{2}\mu_1\cdot B_1\big(\epsilon(\mu_1)\mu_1\big),
\end{eqnarray}
where we used the definitions of $\delta(\mu_1)$, $\epsilon_1(\mu_1)$, and $\epsilon_2(\mu_1)$ in~\raf{parameters}.
By the termination condition of phase I (in line~\ref{s2-0}), we have $B_1(\mu_1)=1$ for $\mu_1\le\delta(m)$. 
We will prove by induction on $\mu_1\ge 1$ that $B_1(\mu_1) \le \mu_1^{\xi(\mu_1)}$. 

Let us assume that $\mu_1>\delta(m)$ and consider first recurrence~\raf{recb1}.  Applying induction we get

\begin{eqnarray*}
	B_1(\mu_1)&\le& \rho\cdot \Big(\Big(1-\frac{\epsilon(\mu_1)}{4(c+1)}\Big)\mu_1\Big)^{\xi(\mu_1)}+1=\rho\cdot \Big(\Big(1-\frac{\ln(2\rho)}{\xi(\mu_1)}\Big)\mu_1\Big)^{\xi(\mu_1)}+1\\
	&\le&\mu_1^{\xi(\mu_1)}\left(\rho e^{-\ln(2\rho)}+\frac{1}{\mu_1^{\xi(\mu_1)}}\right)~~\text{( $\because 1+x\le e^{x}$ for all $x$)}\\
	&<& \mu_1^{\xi(\mu_1)}\left(\frac{1}{2}+\frac{1}{2}\right)=\mu_1^{\xi(\mu_1)}~~\text{( $\because\xi(\mu_1)>1$ for $\mu_1>\delta(m)>2$).}
\end{eqnarray*}
Let us consider next recurrence~\raf{recb2} and apply induction:
\begin{eqnarray*}
	B_1(\mu_1)&\le& \Big(\Big(1-\frac{\epsilon(\mu_1)}{4}\Big)\mu_1\Big)^{\xi(\mu_1)}+\frac{\epsilon(\mu_1)}{2}\mu_1(\epsilon(\mu_1)\mu_1)^{\xi(\mu_1)}
	+1\\
	&=&\mu_1^{\xi(\mu_1)}\left(\left(1-\frac{(c+1)\ln(2\rho)}{\xi(\mu_1)}\right)^{\xi(\mu_1)}+\frac12\mu_1\epsilon(\mu_1)^{\xi(\mu_1)+1}+\frac{1}{\mu_1^{\xi(\mu_1)}}\right)\\
	&\le&\mu_1^{\xi(\mu_1)}\left(e^{-(c+1)\ln(2\rho)}+\frac{1}{4}+\frac{1}{\mu_1^{\xi(\mu_1)}}\right) ~~\text{ ($\because\epsilon(\mu_1)^{\xi(\mu_1)}=\frac{1}{2\mu_1}$ by~\raf{chi})}\\	
		&<&\mu_1^{\xi(\mu_1)}\left(\frac{1}{2\rho}+\frac{1}{4}+\frac{1}{2}\right)\le\mu_1^{\xi(\mu_1)} ~~\text{( $\because\xi(\mu_1)>1$ for $\mu_1\ge\delta(m)>2$ and $\rho\ge2$)}.
\end{eqnarray*}
\qed

\begin{claim}\label{cl4}
	$B_2(\mu_2,t)\le(n\rho)^{2\delta(m^k)\cdot(t+1)}\mu_2^{\xi(\mu_2)}$.
\end{claim}
\proof
There are three possible locations in which a recursive call can be initiated in phase II: 
\medskip

\noindent{\it Line~\ref{s2-11}}: Since $t'\le t-1$, as we remove at least one hypergraph $\cH_i$ by trying all $i$-simple assignments whose number is at most $(|V_0|\rho)^{2|\cH_i|}$, where $|\cH_i|\le \delta(m)\le\delta(m^k)$ for $i=0$ and $|\cH_i|\le \delta(m^k)$ for $i\in[k]$, we get the recurrence   
\begin{equation}\label{recb3}
B_2(\mu_2,t)\le(\rho n)^{2\delta(m^k)}B_{2}(\mu_2,t-1)+1.
\end{equation} 

\medskip

\noindent{\it Line~\ref{s2-13}}: In this case, there are $v\in V_0$ and $i\ne j$ such that $\deg_{\cH_i}(v)\ge\epsilon_1(\mu_2)|\cH_i|$ and $\deg_{\cH_j}(v)\ge\epsilon_1(\mu_2)|\cH_j|$ then the algorithm proceeds similar to lines~\ref{s2-2}-\ref{s2-3}, and we get the recurrence:
\begin{eqnarray}\label{recb4}
B_2(\mu_2,t)&\le&\rho\cdot B_2((1-\epsilon_1(\mu_2))\mu_2,t)+1\nonumber\\
&\le&\rho\cdot B_2\Big(\Big(1-\frac{\epsilon(\mu_2)}{4(c+1)}\Big)\mu_2,t\Big)+1,  
\end{eqnarray}
since we recurse (in the line similar to line~\ref{s2-3}) on a hypergraph $\cH'$ that excludes either all the edges containing $v$ from $\cH_i'$, if we set the color of $v$ to $j$, or all those containing $v$ from $\cH_j'$ if we set the color of $v$ to $i$ (or both, if we set the color of $v$ to $\ell\not\in\{i,j\}$). Note that, in both cases, if $|\cH'|=0$, then $\mu'_2\le\frac{\mu_2}{\delta(m^k)}$ and hence $\mu_2'\le(1-\epsilon_1(\mu_2))\mu_2$ (as $\frac1{\delta(m^k)}<\frac12\le1-\epsilon_1(\mu_2)$).
\medskip

\noindent{\it Lines~\ref{s2-16} and~\ref{s2-19}}: In this case, there is no large-degree vertex. Then Lemma~\ref{l6} implies that there is a $(1-\epsilon_2(\mu_2),1-(\epsilon_2(\mu_2)-(1+c)\epsilon_1(\mu_2)))$-balanced set $S$, with respect to some $\cH_i$, which is found in line~\ref{s2-15}. Then we apply Lemma~\ref{l4} which reduces the problem to one recursive call on the hypergraph $\cH$ in line~\ref{s2-16}, but after fixing the colors of all vertices in $V_0\setminus S$, and at most $\sum_{j\ne i}|\cH_j|\le \mu_2$ recursive calls (in lines~\ref{s2-18}-\ref{s2-19}) on the hypergraphs obtained by fixing the color of one set $H\cap S$, for some $H\in\cH_j$ and $j\ne i$. Note that $S$ satisfies: $(1-\epsilon_2(\mu_2))|\cH_i|\le|(\cH_i)_{S\cup(V\setminus V_0)}|\le(1-(\epsilon_2(\mu_2)-(1+c)\epsilon_1(\mu_2)))|\cH_i|$. In particular, $|\cH_i(V_0\setminus S)|\ge\big((\epsilon_2(\mu_2)-(1+c)\epsilon_1(\mu_2))\big)|\cH_i|$. Since, in line~\ref{s2-16}, we recurse on the hypergraph $\cH'_i:=\cH_i\setminus\cH_i(V_0\setminus S)$, since any $H\in\cH_i$ with $H\cap(V_0\setminus S)\ne\emptyset$ will receive at least one color different from $i$, we get $\mu_2'\le\mu_2(1-(\epsilon_2(\mu_2)-(1+c)\epsilon_1(\mu_2)))$. Moreover, in line~\ref{s2-19}, we will have $|\cH_i'|\leq|\cH_i\setminus(\cH_i)_{S\cup(V\setminus V_0)}|+c$, as (by \raf{intersecting}) all but at most $c$ edges in $(\cH_i)_{S\cup(V\setminus V_0)}$ have non-empty intersections with the set $H\cap S$, in the current iteration of the loop in line~\ref{s2-18}, all vertices of which are assigned the color $j\ne i$. Since $|\cH_i\setminus(\cH_i)_{S\cup(V\setminus V_0)}|\le\epsilon_2(\mu_2)|\cH_i|$, we get the following recurrence for $|\cH_i|>\delta(m^k)\ge\delta(\mu_2)$:

\begin{eqnarray}\label{recb5}
B_2(\mu_2,t)&\le& B_2((1-(\epsilon_2(\mu_2)-(1+c)\epsilon_1(\mu_2))\mu_2,t)+\mu_2\cdot B_2\left(\Big(\epsilon_2(\mu_2)+\frac{c}{\delta(\mu_2)}\Big)\mu_2,t\right)+1\nonumber\\
&\le&B_2\Big(\Big(1-\frac{\epsilon(\mu_2)}{4}\Big)\mu_2,t\Big)+\mu_2\cdot B_2\big(\epsilon(\mu_2)\mu_2,t\big)+1.
\end{eqnarray}
(Note that, if $|\cH'_i|=0$, then $\mu'_2\le\frac{\mu_2}{\delta(m^k)}$ and hence $\mu_2'\le\big(1-\frac{\epsilon(\mu_2)}4\big)\mu_2$, as $\frac1{\delta(m^k)}<\frac12<1-\frac{\epsilon(\mu_2)}{4}$.)

By definition,  $B_{2}(\mu_2,0)=1$, for $\mu_2\ge0$. We will prove by induction on $t=1,\ldots,k$ and $\mu_2\ge1$ that $B_2(\mu_2,t) \le P_{t+1}\mu_2^{\xi(\mu_2)}$, where $P_{t+1}:=\frac{R^{t+1}-1}{R-1}$ and $R:=(\rho n)^{2\delta(m^k)}$. 
We consider 2 cases:

\smallskip

\noindent{Case 1.}~ $1\le\mu_2\le\delta(m^k)$: Then $|\cH_i(\chi)|\le\delta(m^k)$ for all $i\in[0:k]$, and recurrence~\raf{recb3} applies iteratively until we get $t=0$.  By the recurrence, $B_2(\mu_2,1)\le R+1\le P_{2}\mu^{\log_\alpha \mu}$, giving the base case ($t=1$), and by induction on $t$, 
\begin{align*}
B_2(\mu_2,t)&\le R\left(P_t\mu_2^{\log_\alpha \mu_2}\right)+1\le \mu_2^{\log_\alpha \mu_2}\left( RP_t+1\right)= P_{t+1}\mu_2^{\log_\alpha \mu_2}.
\end{align*}

\medskip

\noindent{Case 2.}~ $\mu_2>\delta(m^k)$: 
Let us note first that if the recurrence in~\raf{recb3} applies then the same induction proof (on $t$) in case 1 gives the required bound. Let us note next that  recurrence~\raf{recb4} is identical to \raf{recb1}, but with $B _1(\mu_1)$ replaced by $B_2(\mu_2,t)$ and $\mu_1$ replaced by $\mu_2$. Thus, essentially, the same inductive proof in Claim~\ref{cl3} gives that $B_2(\mu_2,t)\le P_{t+1}\mu_2^{\xi(\mu_2)}$ in this case (as $P_{t+1}>1$).

Finally, let us consider next recurrence~\raf{recb5} and apply induction (on $\mu_2$):
\begin{eqnarray*}
	B_2(\mu_2,t)&\le& P_{t+1}\Big(\Big(1-\frac{\epsilon(\mu_2)}{4}\Big)\mu_2\Big)^{\xi(\mu_2)}+\mu_2P_{t+1}(\epsilon(\mu_2)\mu_2)^{\xi(\mu_2)}
	+1\\
	&=&P_{t+1}\mu_2^{\xi(\mu_2)}\left(\left(1-\frac{(c+1)\ln(2\rho)}{\xi(\mu_2)}\right)^{\xi(\mu_2)}+\mu_2\epsilon(\mu_2)^{\xi(\mu_2)}+\frac{1}{P_{t+1}\mu_2^{\xi(\mu_2)}}\right)\\
	&\le&P_{t+1}\mu_2^{\xi(\mu_2)}\left(e^{-(c+1)\ln(2\rho)}+\frac{1}{2}+\frac{1}{P_{t+1}\mu_2^{\xi(\mu_2)}}\right) ~~\text{ ($\because\epsilon(\mu_1)^{\xi(\mu_2)}=\frac{1}{2\mu_2}$ by~\raf{chi})}\\	
	&\le&P_{t+1}\mu_2^{\xi(\mu_2)}\left(\frac{1}{2\rho}+\frac{1}{2}+\frac{1}{8}\right)<P_{t+1}\mu_2^{\xi(\mu_2)} ~~\text{( $\because\xi(\mu_2)>1$, $P_{t+1}\ge 4$ for $\mu_2\ge\delta(m^k)>2$ and $\rho\ge2$)}.
\end{eqnarray*}
\qed

Using the bounds $\mu_1\le m$,  
$\mu_2\le \delta(m^k)\left(\frac{m}{k}\right)^{k},
$ and $\xi(m^k)\le k\cdot\xi(m)$, we get $B_1(\mu_1)\le m^{\xi(m)}$ and $B_2(\mu_2,k+1)\le(\rho n)^{\frac{2k(k+1)\xi(m)}{\ln(2\rho)}}\left(\frac{\xi(m)}{\ln(2\rho)}\frac{m^{k}}{k^{k-1}}\right)^{k\xi(m)}$. Putting these bounds together, and noting that $\xi(m)\approx\frac{\log_H(2m)}{\log_H\log_H(2m)}$, where $H=4(c+1)\ln(2\rho)$, the lemma follows.
\qed

\begin{lemma}\label{l8}
	The depth of the recursion tree is $O(\frac{h\log m}{\log(2\rho)})$, where $h:=O(\frac{k^2\log_Hm}{\log_H\log_Hm})$ and $H:=4(c+1)\ln(2\rho)$. 
\end{lemma}	
\proof
Let $d_1(\mu_1)$ (resp., $d_2(\mu_2,t)$) denote the depth of the recursion subtree (resp., sub-forest) in phase I (resp., phase II), when the volume of the subproblem is $\mu_1$ (resp., $\mu_2$ and $|\{i\in[0:k]:~|\cH_i(\chi)|\ge1\}|=t$). Then, corresponding to recurrences~\raf{recb1}, \raf{recb2}, \raf{recb3}, \raf{recb4} and \raf{recb5}, we have the following recurrences on the depth: 
\begin{eqnarray}\label{recd1}
d_1(\mu_1)&\le& d_1\Big(\Big(1-\frac{\epsilon(\mu_1)}{4(c+1)}\Big)\mu_1\Big)+1,  \\
\label{recd2}
d_1(\mu_1)&\le& d_1\Big(\max\Big\{\Big(1-\frac{\epsilon(\mu_1)}{4}\Big)\mu_1, \epsilon(\mu_1)\mu_1\Big\}\Big)+1,\\
\label{recd3}
d_2(\mu_2,t)&\le&d_{2}(\mu_2,t-1)+1,\\
\label{recd4}
d_2(\mu_2,t)&\le&d_2\Big(\Big(1-\frac{\epsilon(\mu_2)}{4(c+1)}\Big)\mu_2,t\Big)+1, \\ 
\label{recd5}
d_2(\mu_2,t)&\le& d_2\Big(\max\Big\{\Big(1-\frac{\epsilon(\mu_2)}{4}\Big)\mu_2,\epsilon(\mu_2)\mu_2
\Big\},t\Big)+1.
\end{eqnarray}
Once $\mu_1$ (resp., $\mu_2$) drops to $\delta(m)$ (resp., $\delta(m^k)$), phase I ends (resp., phase II ends after at most $k+1$ more recursive calls). Thus, the above recurrences imply that $d(\mu_1)\le \log_{\alpha} \frac{\mu_1}{\delta(m)}+1$ (resp., $d(\mu_2,k+1)\le \log_{\alpha} \frac{\mu_2}{\delta(m^k)}+k+1$), where $\alpha=\frac{1}{1-\frac{\epsilon(\mu_1)}{4(c+1)}}$ (resp., $\alpha=\frac{1}{1-\frac{\epsilon(\mu_2)}{4(c+1)}}$). It follows that the overall depth of the recursion tree is $O(\frac{k^2\xi(m)\log m}{\log(2\rho)})$.
\qed
\begin{remark}\label{r1}
	If we do not insist on a recursion tree with polylogarithmic depth, then Algorithm~\ref{PKC-alg-B} can be simplified by using $S:=V_0\setminus \{v\}$ for a low-degree vertex $v\in V_0$ in lines~\ref{s2-5} and~\ref{s2-14}. It can be seen from the analysis above that a weaker recurrence will be  obtained with the first term in~\raf{recb2} and~\raf{recb5} replaced by $B_1(\mu_1-\delta)$ and $B_2(\mu_2-\delta,t)$, respectively. The resulting solution will still be $(nm)^{o(\log m)}$ (assuming all other parameters are fixed), but the depth of the recursion tree can be linear in $m$. 
\end{remark}

%\section*{Acknowledgements} 
%The author thanks Endre Boros and Vladimir Gurvich for helpful discussions.
%The author is also grateful for the support received from the Lorentz Center, where this research was initiated. 
%\bibliographystyle{alpha}
%
%\bibliography{dualn}
\newcommand{\etalchar}[1]{$^{#1}$}

%\appendix

%\section{Limiting distribution of Markov chains}
\end{document}